\documentclass{llncs}
\usepackage{indentfirst,amssymb,amsmath,latexsym,graphicx}

\title{Adaptive Cost Model for Query Optimization}

\author{Nikita~Vasilenko, Alexander~Demin, Denis~Ponomaryov}
\institute{Ershov Institute of Informatics Systems, Novosibirsk, Russia \vspace{0.2cm}\newline \email{vasilenko.nikita.research@gmail.com, alexandredemin@yandex.ru, ponom@iis.nsk.su}}

\begin{document}
\maketitle

\begin{abstract}
The principal component of conventional database query optimizers is a cost model that is used to estimate expected performance of query plans. The accuracy of the cost model has direct impact on the optimality of execution plans selected by the optimizer and thus, on the resulting query latency. Several common parameters of cost models in modern DBMS are related to the performance of CPU and I/O and are typically set by a database administrator upon system tuning. However these performance characteristics are not stable and therefore, a single point estimation may not suffice for all DB load regimes. In this paper, we propose an Adaptive Cost Model (ACM) which dynamically optimizes CPU- and I/O-related plan cost parameters at DB runtime. By continuously monitoring query execution statistics and the state of DB buffer cache ACM adjusts cost parameters without the need for manual intervention from a database administrator. This allows for responding to changes in the workload and system performance ensuring more optimal query execution plans. We describe the main ideas in the implementation of ACM and report on a preliminary experimental evaluation showing 20\% end-to-end latency improvement on TPC-H benchmark.
\end{abstract}


\section{Introduction}
The convenience of (declarative) query languages goes along with the complexity of finding an optimal execution plan for a query. It is well-known that for a SQL query there may exist an exponential number of execution plans (in the size of the query) and the problem to find one with a minimal latency is NP-complete. In the search for near-optimal execution plans, a typical query optimizer relies on a cost model which provides an estimation of the expected plan latency in terms of an abstract cost. In conventional implementations, a cost model is essentially a collection of formulas providing integer costs to the physical implementation of query operators and formulas to aggregate these costs into the resulting cost of a plan. The accuracy of the cost model has direct impact on the optimality of plans selected by the optimizer and therefore, on the end-to-end query latency and overall DB performance. A higher cost would typically mean a higher execution time for a plan, but in practice these values are often decorrelated, non-optimal plans are selected by the optimizer and some queries appear to be slow.  
   
A typical cost-based query optimizer relies on three main components: cardinality estimation, CPU and I/O cost estimation, and plan enumeration. Cardinality estimation means predicting the number of records in the output of a particular plan node. Numerous techniques to improve the accuracy of cardinality estimation have been proposed in the literature including \cite{2020arXiv200909884H,2018arXiv180900677K,sel_est,10.1145/3318464.3389741}. The component related to CPU and I/O cost estimation is also important for accurate estimation of query plan performance, because performance of a plan (when seen as a program) is obviously affected by CPU and I/O load at the execution time \cite{2021arXiv210101507L,6544899}.  Finally, plan enumeration is to find an execution plan with the lowest cost.  By employing all these components, the task of an optimizer is to find the most efficient execution plan for a query, thereby improving the overall DB performance. However this approach has a significant drawback: in modern DBs cost estimation may be wrong by \textbf{several orders of magnitude} (e.g., when estimating cost of complex queries with multiple joins \cite{6544899}). 

Recently, an idea of learning-based optimizer leveraging prior query execution experience has been proposed \cite{10.14778/3342263.3342644,10.14778/3583140.3583160,6228100}. This approach tries to mitigate the issues of conventional cost-based optimizers by employing learning from complex prior examples. The downside of this method is that it is well-suited for static workloads, but in the case of data drift, changes in query patterns, or data schema updates the optimizer has to be retrained.

In \cite{10.1007/s00778-017-0480-7}, the authors showed that to reduce cost estimation errors, one has to improve both, cardinality estimation accuracy and cost model parameter tuning. In this work, we argue that dynamically tuning CPU- and I/O-related cost model parameters alone can improve optimizer decisions in the search for an optimal plan.

Typically, a database administrator configures these cost model parameters manually. Finding the most suitable values for these parameters requires a deep understanding of the database structure, workload characteristics, and system resources. Inaccurate parameter setting can lead to suboptimal execution plans and degraded performance. Furthermore, cost parameters may need to be periodically adjusted to adapt to changes in workload, data access patterns, and system performance. This makes selecting optimal cost parameters complicated.

Another approach is to execute queries from a specifically designed calibration workload and collect execution statistics on it \cite{6544899}. Then cost model parameters can be adjusted based on these statistics. However in this approach one has to periodically execute a synthetic calibration workload in addition to the real DB workload and even in this case the obtained parameter estimations may be inadequate for the real workload due differences with calibration queries. 

In this work, we propose an \textbf{adaptive cost model (ACM)}, a novel approach for optimizing CPU- and I/O-related plan cost parameters which implements lightweight ML models to analyze statistics on executed queries and makes prediction on the parameter values for next queries. The solution consists of two components: a model for estimating cost parameters related to CPU performance and a model for estimating disk-related parameters. The approach implements online fine-tuning which allows for dynamically adapting to workload changes. It avoids the need of using calibration queries, thereby reducing the overall load on database and also on DBAs by improving query plans and reducing the need to manually tune a number of configuration parameters. 

We begin our exposition in Section \ref{Sect:Background} with a brief summary on related work and the main ingredients of cost models in conventional database optimizers. Then we describe in Section \ref{Sect:Main} the main ideas and design solutions in Adaptive Cost Model: the CPU- and disk-related components. Finally, in \ref{Sect:Experiments} we provide results of a preliminary experimental evaluation on TPC-H benchmark and in Section \ref{Sect:Conclusion} we conclude.

\section{Background}\label{Sect:Background}
\subsection{Related Work}
In the literature, there is a number of approaches to configuring grand unified configuration parameters (GUCs) of databases, including those related to cost model. The most general methods rely on search techniques like random search, simulated annealing, and Reinforcement Learning \cite{2021arXiv211210925T,10.1145/3299869.3300085,10.14778/3352063.3352129,10.1145/3035918.3064029,10.5555/3488733.3488749}, to name a few. In these approaches, cost model parameters are treated as general database configuration parameters and it is suggested to optimize them with search techniques just like any other DB parameters. 

In \cite{2021arXiv211210925T}, the authors proposed to use a language model combined with Reinforcement Learning techniques to identify DB configuration parameters from manuals and posts in open DBA discussion groups. In \cite{10.1145/3299869.3300085}, a deep deterministic policy gradient method was proposed to find optimal configurations in high-dimensional continuous spaces with Reinforcement Learning.

In \cite{10.14778/3352063.3352129}, the authors propose to consider three levels of tuning with Reinforcement Learning. At the query level, GUCs are tuned separately for each query prior to execution by representing a query as a vector and recommending a configuration. At the workload level, the configuration is tuned for the entire workload by combining all query vectors into one and generating a recommended configuration. At the cluster level, queries are separated into clusters and configurations are tuned for each cluster.

In \cite{10.5555/3488733.3488749}, the authors identify the most important GUCs for tuning by first using Latin Hypercube Sampling (LHS) to sample configurations for all GUCs. They then collect statistics for these samples by running benchmarks for each configuration. Then by using a random forest, they rank  GUCs to determine the most significant ones.

In the approach presented in \cite{10.1145/3035918.3064029}, the authors propose to begin with selecting the closest matching configuration for a database. During an initial observation period, they characterize the workload and hardware by using internal runtime metrics and by running queries. Factor analysis and K-means clustering are used to reduce the value of the metric and the result is used for system profiling. The most important knobs are selected by using Lasso and ranking. Tuning proceeds incrementally, increasing the number of knobs during the process. Initially, they find the closest similar workload from the repository, and then they determine the best configuration by using a Gaussian Process.

Another class of approaches is based on the idea of using calibration queries and formulas to adjust cost parameters. For example, in \cite{6544899}, based on the observed latency of calibration queries, the values of the parameters are derived analytically from a set of formulas. 

We can summarize all these approaches as follows:
\begin{itemize}
    \item Search-based approaches:
        \begin{itemize}
            \item provide more accurate parameter values
            \item are time-costly
        \end{itemize}
    \item Calibration-based approaches:
        \begin{itemize}
            \item allow for quickly estimating parameter values
            \item face the problem of constructing a calibration query for each parameter
            \item may provide relatively high errors preventing accurate calculation of optimal values
        \end{itemize}
\end{itemize}
Both approaches share the common disadvantage: they require a large number of trials to assess the impact of a set of parameter values on workload execution time.

\subsection{Cost Model in DB Optimizers}
Conventional open-source cost-based databases with bottom-up optimization (e.g., relational DBs including PostgreSQL, MySQL, openGauss, etc.) employ a cost model which includes the following five key parameters:
\begin{enumerate}
    \item \(cpu\_tuple\_cost\), the CPU cost to process a tuple,
    \item \(cpu\_index\_tuple\_cost\), the CPU cost to process a tuple via index access,
    \item \(cpu\_operator\_cost\), the CPU cost to perform an operation such as hashing or aggregation,
    \item \(seq\_page\_cost\), the I/O cost to sequentially access a page,
    \item \(random\_page\_cost\), the I/O cost to randomly access a page.
\end{enumerate}

The cost of the query execution plan is calculated as the sum of the costs of the operators in the plan. The cost of operators is calculated as a linear combination of basic parameters (cost-relevant GUC parameters being numeric constants) and variable values such as the number of data blocks read from disk, etc.

In general, the cost of a SQL operator is computed as the sum of the products of cost model parameters and estimated cardinalities:
\begin{equation}
    Cost_{operator}=c_t \cdot n_t+c_o \cdot n_o+c_s \cdot n_s+c_i \cdot n_i+c_r \cdot n_r,
\end{equation}
where:
\begin{itemize}
    \item \(c_t\) - cost of processing a tuple,
    \item \(n_t\) - number of tuples processed,
    \item \(c_o\) - cost of performing an operation,
    \item \(n_o\) - number of operations performed,
    \item \(c_s\) - cost of a sequential disk page fetching,
    \item \(n_s\) - number of disk pages fetched sequentially,
    \item \(c_i\) - cost of processing an index entry,
    \item \(n_i\) - number of index entries processed,
    \item \(c_r\) - cost of fetching a disk page randomly,
    \item \(n_r\) - number of disk pages fetched randomly.
\end{itemize}

For example, the cost of sequential access to a table (SeqScan operator) without filter can be defined as:

\[Cost_{SeqScan}=c_t \cdot n_t+c_s \cdot n_s\]

and the cost of aggregation operator is:
\[Cost_{Agg}=c_t \cdot n_t\]

The optimal choice of values for the parameters above depends on the hardware configuration, workload type, as well as on the ‘warm-up degree’ of the database, i.e., the amount of data which resides in the buffer cache.

In the above mentioned list of parameters, \(seq\_page\_cost\) is typically set to  the time required to process one page in sequential scan. The rest of the parameters are set as fractions of this base value. For instance, \(cpu\_tuple\_cost\) set to 0.01 means that processing 100 rows requires the same time as reading one page sequentially from disk. %\(cpu\_index\_tuple\_cost\) equal to 0.005, and \(cpu\_operator\_cost\) equal to 0.0025. 
Similarly, \(random\_page\_cost\) equal to 4 means that the time required for random access to a page exceeds the sequential access time by a factor of 4.

Thus, to increase the cost of sequential page scan, one can decrease the values of the other four parameters, and vice versa, to increase the cost of all operations except \(seq\_page\_cost\), it suffices to decrease only \(seq\_page\_cost\). Therefore, to adjust disk and CPU-related configuration of a cost model it suffices to change only four parameters. W.l.o.g. we assume in this paper that \(seq\_page\_cost\) is 1.

The primary objective of cost model parameter tuning is to increase the correlation between the cost and the execution time of a node, which allows for  estimating the execution time of plans more accurately. Thus, the task looks like:
\begin{equation}
    corr(cost, time) \rightarrow 1
\end{equation}
where
\begin{itemize}
    \item \(cost\)  is a cost for a plan operator,
    \item \(time\) is a real execution time for this operator.
\end{itemize}

\section{Design of ACM}\label{Sect:Main}

ACM implements computation of the cost parameters associated with disk I/O operations (I/O cost) and cost parameters related to the processor utilization (CPU cost) in two different ways. Having two different approaches is natural, because CPU parameters do not depend on the current state of the database, while disk-related parameters can change depending on the state of the database buffer cache, which varies nearly after every query.

ACM computes optimal I/O cost parameters before every query by assessing the volume of requested data already present in the database buffer cache and by estimating the amount of data that needs to be retrieved from the disk. The model employs historical data on the execution of past queries to generate accurate predictions. 

To calculate optimal CPU-related cost parameters, the model is continuously monitoring query operator execution statistics and computes parameters to align the estimated execution time (determined at the planning stage) as close as possible to the actual execution time (obtained post query execution). This way ACM continuously adjusts the cost model parameters to provide more accurate estimates of plan execution costs, which guides the database optimizer towards more optimal query plans.

In the following, we provide a detailed description of the two components of ACM: the disk-related model and CPU-related model.

\subsection{Disk-Relevant Parameters}

Heuristic-based formulas of the conventional cost model take into account the difference between the cost of sequential and random page access by differences in cost parameters. However, this approach does not allow for estimating the number of requested pages that are already in the buffer cache and thus, do not have to be fetched from disk. Since the optimizer is not aware of how much of the requested data already resides in the memory, it makes estimation errors. ACM improves the accuracy of cost estimation by dynamically adjusting the cost of random disk access based on the information about the availability of data in the buffer.

When a query is executed, ACM collects statistics on the usage of data from the buffer (buffer hit) and disk (read). This relationship is important because it gives an idea of how much of the requested data was in the database memory.
First of all, the proposed model separately calculates the \(random\_page\_cost\) parameter for each table, as one table might reside entirely in the buffer cache, while another table might be absent. Having the entire table in the buffer cache allows for setting \(random\_page\_cost\) equal to \(seq\_page\_cost\), suggesting to the optimizer that the read time will be the same. Thus, ACM implements a dynamic computation of \(random\_page\_cost\)  relative to tables. As already mentioned, an exact estimate for \textit{random\_page\_ cost}  depends on the information on how much of the requested data will be taken from the buffer and how much will be read from disk. This information can be expressed by the 'hit ratio' defined as:
\begin{equation}
    R(q) = \frac{hit}{hit + read}
\end{equation}
where
\begin{itemize}
    \item \(q\) is a query,
    \item \(hit\) is the number of pages that was requested by the query and read from the memory,
    \item \(read\) is the number of pages that was requested by the query and read from the disk.
\end{itemize}

Typically, this information becomes available after query execution, but ACM tries to estimate this ratio by using statistics on the execution of previous queries. For each table, the model collects hit ratio statistics:
\begin{equation}
    R(q_{t}), R(q_{t-1}), R(q_{t-2}), ...
\end{equation}
where
\begin{itemize}
    \item \(R(q_{t})\) is the hit ratio of query \(q_{t}\) executed at time \(t\).
\end{itemize}

To estimate the hit ratio \(R(q_{t+1})\) for a new query \(q_{t+1}\) one can use various models to predict the hit ratio based on previous values \(R(q_{t})\), \(R(q_{t-1})\), ... . For example, one can use various time series prediction models like ARIMA, exponential smoothing, etc.. But in ACM we employ a simpler solution which analyses only the previous hit ratio and a degradation factor. In fact, in our experiments, this simpler model showed results as good as more complex ones:
\begin{equation}
    R(q_{t+1}) = R(q_{t}) \cdot D ,
\end{equation}
where $D$ is a degradation factor defined as:
\begin{equation}
    D = \frac{1 + (qc - tc)}{1 + (qc - tc)^2} 
\end{equation}
and
\begin{itemize}
    \item \(qc\) is a query counter (current total number of accesses to all tables; it gets incremented upon any table read),
    \item \(tc\) is a counter that indicates the last access to the target table. If target table is accessed, the current value of \(qc\) is written to \(tc\).    
\end{itemize}

The meaning of the degradation factor is explained as follows. In general, after a query referring to a table \(T_{1}\) is executed, some other queries referring to tables \(T_{2}, ..., T_{n}\) may be processed, which may cause the data of table \(T_{1}\) to be evicted from the buffer. The more requests to  different tables have been executed since the last access to table \(T_{1}\), the less likely its data will remain in the buffer. The degradation factor takes this effect into account.

Next, it is necessary to relate \(R(q_{t+1})\) to \(random\_page\_cost\) so that at maximum the predicted hit ratio \(random\_page\_cost\) equals \(seq\_page\_cost\). We set:
\begin{equation}
    \begin{split}
        random\_page\_cost_{new} = random\_page\_cost_{def} \cdot \\ (1 - R(q_{t+1})) +  seq\_page\_cost \cdot R(q_{t+1})
    \end{split}
\end{equation}
where
\begin{itemize}
    \item \(random\_page\_cost_{def}\) is the default value of \(random\_page\_cost\),
    \item \(random\_page\_cost_{new}\) is the new value for \(random\_page\_cost\) that is used to calculate costs of scans for a particular table.
\end{itemize}

\begin{figure}[h]
    \centering
    \includegraphics[scale=0.45]{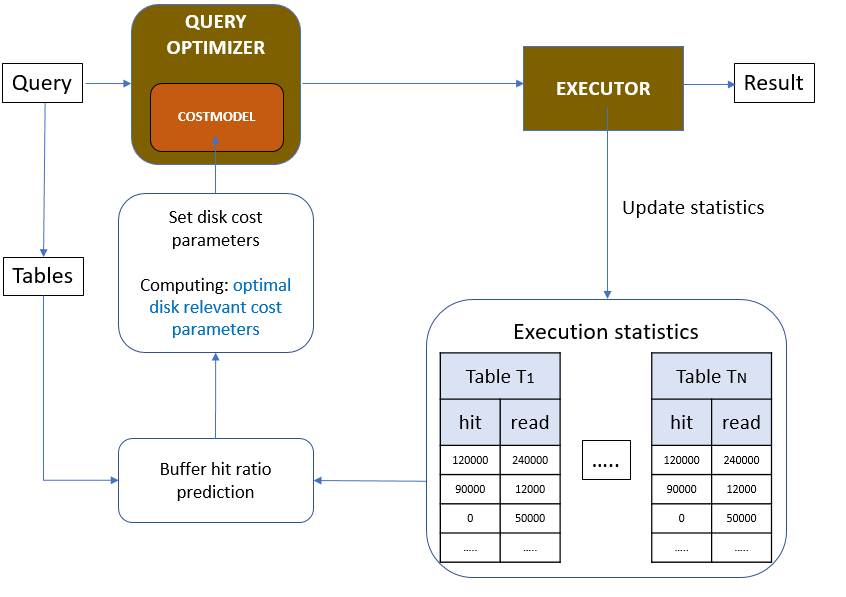}
    \caption{Computation of disk-relevant cost parameters}
    \label{fig:disk architecture}
\end{figure}

Figure \ref{fig:disk architecture} shows the schema of computing buffer usage statistics at query execution. The more pages are read for the table from the memory, the lower the cost of subsequent random accesses to the memory will be, since most pages can be read from the buffer. Thus, the cost of random access to the page is adjusted for each data table.

The number of pages read from disk and memory is collected for each table until the amount of data reaches the level when it is possible to predict the hit ratio for a next table access. After that, at query planning, the cost of random access to the page is computed by using the degradation factor as defined above and the obtained costs are injected. ACM considers the size of the time interval between the last access to the table and the scheduled one, changing the number of hits to the degradation factor. With cost injection, query planning is standard and after a query is executed new parameter values will be recorded for all the tables with the data.

\subsection{CPU-Relevant Parameters}
\begin{figure}[h]
    \centering
    \includegraphics[scale=0.4]{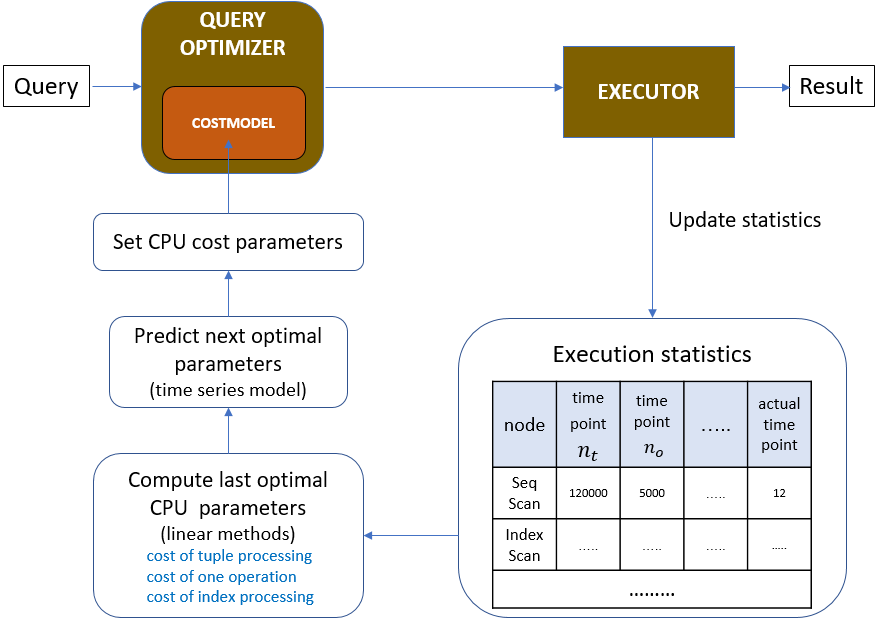}
    \caption{Computation of CPU-relevant cost parameters}
    \label{fig:cpu architecture}
\end{figure}

The other three parameters in the standard cost model refer to CPU operations, such as the cost of processing one tuple (tuple cost), performing one operation (operator cost), and processing one index tuple (index tuple cost). Standard parameter values do not reflect specifics of various hardware setups and database configurations which again may lead to inaccurate estimates of query costs. Also, standard cost models do not provide individual cost computation formulas for different operators which is another source of inaccuracy. To address these issues, ACM provides a method to collecting statistics and dynamically set parameters to account for a DB configuration and hardware setup individually for each plan operator. Figure \ref{fig:cpu architecture} shows the logic of statistics collection on CPU usage at query execution.

After a query is executed, the system collects key statistics including the execution time for operators and the number of tuples processed by each operator. Statistics for each type of operator are collected separately. The purpose of using these statistics is to reduce the mismatch between the output values of operator cost formulas and the execution times for these operators. The minimization function uses linear lightweight models, which have a minimal computation overhead. An operator cost is defined as:
\begin{equation}
    \begin{split}
        & F(c_t, c_o, c_i) = c_t \cdot n_t+c_o \cdot n_o + c_i \cdot n_i + s, \\
        & s = c_s \cdot n_s + c_r \cdot n_r
    \end{split}
    \label{eq:to_min}
\end{equation}
where
\begin{itemize}
    \item \(c_t\) is a cost of processing a tuple,
    \item \(n_t\) a number of tuples processed,
    \item \(c_o\) a cost of performing an operation,
    \item \(n_o\) a number of operations performed,
    \item \(c_i\) a cost of processing an index entry,
    \item \(n_i\) a number of index entries,
    \item \(s\) a cost of sequential and random reads for a node with GUC parameters derived from disk-relevant model.
\end{itemize}

Based on this cost formula, a simple linear model is defined:
\begin{equation}
    L(c,x) = x^T c
    \label{eq:model}
\end{equation}
where the vector \(c=(c_t, c_o, c_i, 1)\) of model parameters and \(x=(n_t, n_o, n_i, s)\) are independent variables. Note that \(s\) is treated as an independent variable because its value is set by the disk-relevant model.

Suppose we have a sample of observations \(X\) of variable values collected by the system at query execution:
\begin{equation}
    X=
    \begin{pmatrix}
        n_t^1 & n_o^1 & n_i^1 & s^1\\
        n_t^2 & n_o^2 & n_i^2 & s^2\\
        ... & ... & ... & ... \\
        n_t^n & n_o^n & n_i^n & s^n\\
    \end{pmatrix}
\end{equation}
We can rewrite the model \ref{eq:model} in the matrix form:
\begin{equation}
    L(c,X) = Xc
    \label{eq:model_matrix}
\end{equation}

and then the task of finding optimal cost parameters is reduced to the linear regression problem:
\begin{equation}
    L(c,X) \rightarrow time \cdot scale\_factor
\end{equation}
where 
\begin{itemize}
    \item 
    \(time=
    \begin{pmatrix}
        t^1\\
        t^2\\
        ...\\
        t^n\\
    \end{pmatrix}
    \) is a collected execution time for an operator,
    \item \(scale\_factor\) is a  constant representing a scaling factor between execution time and cost.
\end{itemize}
Various techniques can be used here to estimate model parameters. In our implementation we used the classical least squares method.

After collecting statistics and calculating cost parameters, ACM saves the results for predicting cost parameters for future queries. When adjusting parameters dynamically, it is important to respond in a timely manner to changes in database configuration, as well as to other factors that may affect cost. ACM uses time series analysis to estimate future cost parameters when they are individually set for each operator type.
In our implementation we used the exponential moving average, but we note that other models (e.g. ARIMA, etc.) can be used. The exponential moving average is defined as:

\begin{equation}
    c_{pred, n} = (1 - \alpha) \cdot c_{n-1} + \alpha \cdot c_{pred, n - 1}
\end{equation}
where 
\begin{itemize}
    \item \(c_{pred, i}\) is a predicted value for cost parameter on step \(i\),
    \item \(c_i\) a cost parameter value derived from the linear model (\ref{eq:model_matrix}) on step \(i\),
    \item \(\alpha\) a smoothing factor.
\end{itemize}

In this way, ACM employs query execution statistics to dynamically adjust 5 essential cost model parameters, which enables the DB optimizer to estimate cost more accurately for each type of operator. Information analyzed by ACM includes (but is not limited to) buffer state information, execution time of queries and nodes, the number of tuples produced and other key feature values. 

\section{Preliminary Experimental Evaluation}\label{Sect:Experiments}
We now present first results of an experimental evaluation of ACM on the standard TPC-H benchmark \cite{tpch} in which we answer the following questions:
\begin{itemize}
\item Is correlation between time and cost for nodes increased by ACM?
\item Does ACM improve benchmark execution latency? 
\end{itemize}
%We answer these questions, we need to consider not only the end-to-end execution time results but also to show the correlation magnitude between time and cost for nodes.

For experiments, we used a 10Gb TPC-H dataset consisting of 22 template-based queries. We implemented ACM in the open-source database openGauss, which has a cost model similar to PostgreSQL. The benchmark was run several times to `warm up'  buffer followed by the execution of test queries with measurements.

The correlation between cost and execution time of plans for TPC-H queries without/with ACM is shown in Figures 3. One can notice two groups of nodes on the left part of the figure: one group has time within 200 ms and cost value up to 5000, while the second group lies significantly higher, with costs ranging from 20,000 to 25,000 and times from 200 to 400 ms. The correlation between cost and time is 0.29.
\medskip

\begin{tabular}{cc}
\includegraphics[scale=0.3]{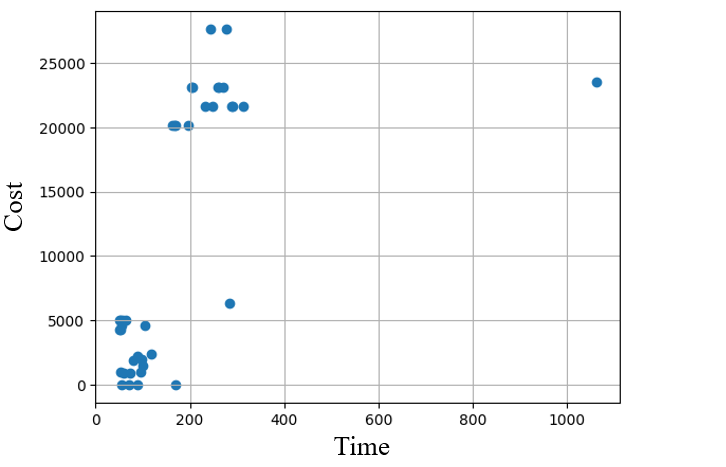} & \includegraphics[scale=0.3]{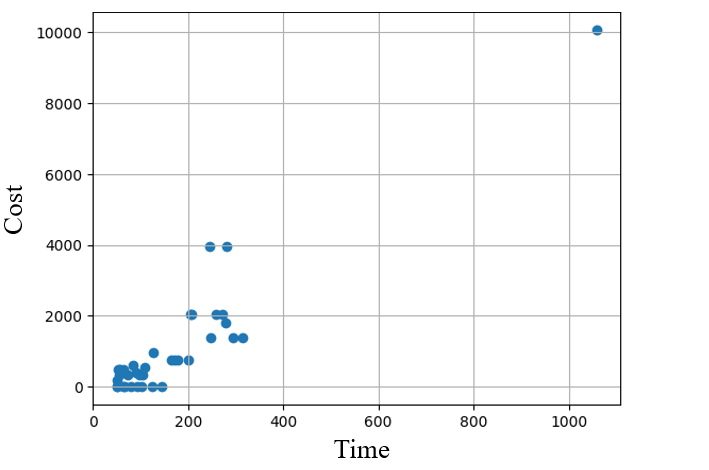}
\end{tabular}

\begin{center}
{\footnotesize \textbf{Fig. 3.} Correlation between node cost and execution time: with standard openGauss cost model (left) and with ACM(right)}
\end{center}
%\begin{figure}[h]
 %   \centering
   % \includegraphics[scale=0.3]{corr before.PNG}
    %\caption{Correlation between node cost and execution time with standard openGauss cost model}
    %\label{fig:corr_before}
%\end{figure}

When using ACM, the correlation is increased to 0.92, which means a big improvement of the cost model accuracy.

%\begin{figure}[h]
 %   \centering
   % \includegraphics[scale=0.3]{corr after.PNG}
    %\caption{Correlation between node cost and execution time with ACM}
    %\label{fig:corr_after}
%\end{figure}
 \setcounter{figure}{3}
Figure \ref{fig:tpch_results} shows end-to-end latency of TPC-H queries. 

\begin{figure}[h]
    \centering
    \includegraphics[scale=0.3]{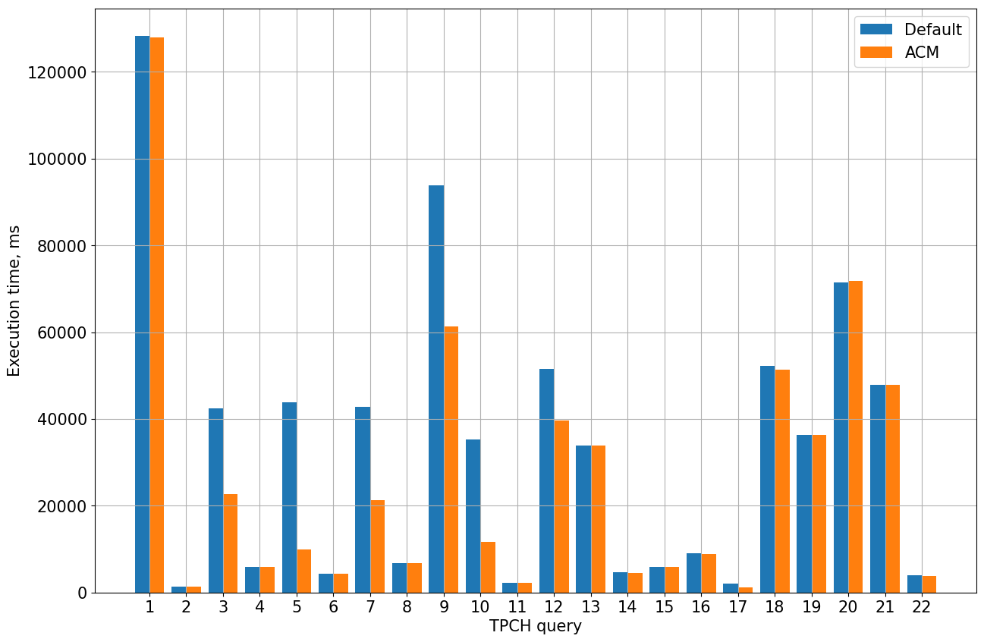}
    \caption{E2E latency of TPC-H queries: summary}
    \label{fig:tpch_results}
\end{figure}

The query plan is changed for queries 3, 5, 7, 9, 10, 12, and 17. For the remaining queries, the execution plan remains unchanged. The overall latency improvement for queries with a modified plan is 46\% and the latency improvement for the entire benchmark is 20\%. We provide detailed latency results (in milliseconds) in Table \ref{tab:tpch_full_results}.

%\begin{center}
\begin{table}\caption{E2E latency values for TPC-H queries}\label{tab:tpch_full_results}
\centering
\begin{tabular}{c|c|c||c|c|c}
Query No. & Baseline & ACM & Query No. & Baseline & ACM  \\ \hline
1 & 128235 & 127906 & 12 & 51593 & 39639 \\ \hline
2 & 1325 & 1423 & 13 & 33872 & 33795  \\ \hline
3 & 42453 & 22632 & 14 & 4685 & 4442 \\ \hline
4 & 5969 & 5956 & 15 & 5893 & 5949 \\ \hline
5 & 43756 & 9860 & 16 & 8992 & 8816 \\ \hline
6 & 4290 & 4254 & 17 & 1971 & 1225 \\ \hline
7 & 42803 & 21306 & 18 & 52151 & 51435 \\ \hline
8 & 6819 & 6808 & 19 & 36333 & 36371 \\ \hline
9 & 93829 & 61407 & 20 & 71459 & 71743 \\ \hline
10 & 35248 & 11726 & 21 & 47928 & 47944 \\ \hline
11 & 2308 & 2225 & 22 & 3909 & 3878 \\ \hline
\end{tabular}

\end{table}
%\end{center}

We  note that some queries could not be sped-up with ACM which was primarily caused by errors in cardinality estimation. Even though ACM improves key parameters of the cost model it may not be able to shift the choice of optimizer towards better plans in all cases due to inaccuracy of cardinality estimation.

\bigskip

\section{Conclusion}\label{Sect:Conclusion}
In this paper, we have proposed a method to dynamically adjust CPU- and I/O-related parameters of a conventional cost model of a database optimizer. Our approach differs from other solutions known in the literature in that it is online, it does not require pre-training or periodic running of calibration queries on the database, and it is based on light-weight learning models with a minimal computational overhead, which makes it  practical for implementation in DB kernels. The proposed solution takes into account several nuances for a more accurate cost estimation including buffer hit ratio prediction for accessed tables and per operator cost estimation. Even though improving the accuracy of CPU- and disk-related cost model parameters alone can not improve the optimality of plans in all cases, we have demonstrated in preliminary experiments on TPC-H benchmark that it may have a significant positive impact on query latency due to improved correlation between cost and execution time. The next natural step would be to evaluate this approach in conjunction with methods to improve cardinality estimation and planning in general, and we leave this for future work.  

%We described ACM, a tool for real-time tuning of cost model parameters based on statistics collected from query execution results. ACM comprises two components: a CPU-related model, which adjusts the cost model parameters related to CPU performance, and a disk-related model, which utilizes buffer cache usage information to predict cost model parameters associated with page reads from memory and disk. Testing conducted on the TPC-H benchmark demonstrated the importance of tuning cost model parameters and the necessity of considering buffer states for accurate plan evaluation. 

\bibliographystyle{acm}
\bibliography{mybibliography}

\end{document}